# Biospeckle signal descriptors: a performance comparison

Ana L. Dai Pra[a], Isabel L. Passoni[b], G. Hernan Sendra [c]∗
Marcelo Trivi[d], Hector. J. Rabal[d]

[a] *Grupo Inteligencia Artificial, Departamento de Matemática.*
[b] *Laboratorio Bioingeniería. Departamento de Electrónica*
*Facultad de Ingeniería. Universidad Nacional de Mar del Plata, Argentina.*
*4302 Juan B. Justo Av., Mar del Plata, Bs. As., Argentina; B7608FDQ;*

[c] *Imaging Facility, Zentrum für Molekulare Biologie der Universität Heidelberg. Im Neuenheimer Feld 282, D-69221 Heidelberg, Germany. E-mail:gh.sendra@zmbh.uni-heidelberg.de*

[d] *UID Optimo, Departamento de Ciencias Básicas. Facultad de Ingeniería.*
*Universidad Nacional de la Plata, Argentina.*

**Abstract**

The characterization of dynamic phenomena using laser speckle has been studied in several applications and many descriptors have been developed to analyze the resulting images. In this work we compare the performance of a set of dynamic speckle descriptors by applying them to carefully controlled numerical simulations in order to explore their linearity, robustness, sensitivity related to the samples quantity, as well as also by their computing time. Results are shown as plots together with discussions about the corresponding advantages and shortcomings. Also the robustness to inhomogeneous spatial intensity was evaluated in an experiment performed with the illuminated surface of an actual biological object

1. Introduction

When an object surface, illuminated by a coherent beam, presents some type of local movement, the intensity and shape of the observed speckles evolve with time. The speckle patterns thus become time dependent. This phenomenon is also characteristic of biological samples and is known as biospeckle. The biospeckle activity is the consequence of microscopic movements or local changes in the refractive index of the sample properties.

Biospeckle patterns can be classified also as "boiling" patterns [1] the speckles move, deform, disappear and reappear without any significant displacement of their mean position. This behavior can also be observed in some non-biological processes, such as drying of paints, corrosion, etc.

Both the time evolution of a pixel intensity and its spatial distribution over an image show seemingly random variations similar to those found in the height distributions of a rough surface. The characterization of rough surfaces requires the measurement of a considerable set of parameters. A similar behavior can be expected for speckle patterns, added to the fact that the inner dynamics of the process that produces them are most of times poorly known. Therefore many descriptors found in the literature are either heuristic or describe only some of the mechanisms.

Biological samples include many variables and show inherent variability; hence different contributions to dynamic speckle cannot easily be assigned

Although many descriptors have been developed to characterize dynamic speckle, not all of them are suitable for all applications and there is no well defined criterion on how to make a choice. This work presents some approaches to analyze, compare, evaluate and select descriptors on simulated series where it is possible to assess the awaited behavior when studying an application. A similar comparison has been reported with respect to the drying of paint measuring [2]. The effect of using a comparatively small number of samples was explored in [3].

With this aim a set of simulated speckle patterns was generated by using a numerical model that had been tested with actual experiments [4]. It is based on the simulation of an in-plane moving diffuser in pure boiling conditions; where the activity level varies according to the speed of the diffuser in a controlled way.

Several descriptors were analyzed: Standard deviation, Temporal contrast, Full width at half maximum of autocorrelation, High to low ratio, Cutoff frequency, Shannon entropy, Wavelet entropy, Energy of spectral band, Averaged Differences, Generalized Differences, Weighted generalized differences, Subtraction average and Fuzzy granular analysis.

The evaluation was done in terms of linearity with respect to the diffuser velocity, convergence and performance against the required quantity of images required. The computational time was also compared among the descriptors. Finally, the robustness to intensity variations within the sample was evaluated in a real and complex experiment where a non-visible bruise in apples was detected.

## 2. Methods

2.1 Controlled simulations

In order to evaluate the performance of different algorithms, a huge set of samples is required to ensure an acceptable statistical representation of certain situation. Experimentally obtained Biospeckle samples are subject to inherent variability; in addition, repeatable samples require controlled environmental conditions. Alternatively, numerical simulations obtained with a model of generation of dynamic speckle can provide a better approach to analyze the performance of descriptors. There are several existing numerical models to simulate speckle patterns [5][6][7][8], from where we have chosen the model by Sendra et al., whose numerical model and required conditions for boiling observation are described in [4]. It adequately describes a simple and repeatable boiling experiment, namely the speckle pattern produced by an in-plane moving diffuser under adequate geometrical conditions of the observation system. This numerical experiment can be reproduced in practice [9].

When a speckle pattern is generated by the motion of a rigid diffuser under the most general conditions it is possible to observe two phenomena. The first is the translation of the pattern as a whole and consequently also the translation of each speckle grain. The other appears when grains speckle change their shape and eventually vanish or are created as the diffuser moves and is called *boiling*. Generally, both phenomena occur simultaneously; however, there are conditions for which it is possible to observe patterns of pure translation or pure boiling. Okamoto and Asakura [1] describe these conditions for the case of fully developed speckle patterns, in which the scattering centers in the diffuser are uncorrelated with each other and the diffuser moves with certain speed. They use the normalized autocorrelation of the space-temporal intensity function. The samples used in this work were numerically simulated in pure boiling condition.

2.2. Actual experiment: bruise-apple

To detect bruised apple regions a sequence of 500 whole field 300x300 speckles images from red delicious apples was obtained. The damage was caused by a controlled impact produced by letting fall a steel ball on the apple, which cannot be appreciated by visual inspection. The images were assembled into a three-dimensional array, hence 90,000 (300x300) series are processed [10]. Then, an image of the sample is constructed with the descriptor value of each series. The process enables the segmenting of regions with different bio-activity levels, essentially bruised and non-bruised zones. The speckled images were not obtained by free propagation but consisted in subjective speckle focused images formed by an objective (usually f = 50mm, f/# = 16). An expanded laser beam is used for the illumination and the CCD camera registers the image with Fraunhofer subjective speckle.
.

## 3. Descriptors

There are several descriptors defined in the literature. A sub-set of them was considered for this work and are explained in this section. In the following, the variable *X* will be considered as the signal that represents the temporal evolution of the intensity of a point in the speckle pattern, with a quantity of samples given by *N*. Hence, $x_i$ is the *i*-th individual element of the *X* time series, where *i*: 0, 1, … *N*.

Three categories of descriptors are proposed as they are based on: statistical analysis, processing in the frequency domain and processing in the time domain,

3.1 Descriptors based on statistical analysis

*3.1.1 Standard deviation (SD)*

From the statistical point of view, the simplest method to detect variations in a signal is the standard deviation (*SD*), which is a measurement of the spread of the variations in the signal [11].

$$SD = \sigma(X) \quad (1)$$

*3.1.2 Temporal Contrast (TC)*

In actual experiments, the standard deviation is related and thus very sensible to the mean intensity of the speckle pattern, since the speckle is not a linear process. A first way to solve this problem is the division of the standard deviation by the mean intensity. In order to differentiate this descriptor from the spatial image contrast used by Briers [12], we have called it "temporal contrast".

$$TC = \frac{\sigma(X)}{\langle X \rangle} \quad (2)$$

where <.> means temporal average.

*3.1.3 Full Width Half Maximum of the Autocorrelation (FWHMA)*

The first order statistics have the disadvantage of not taking into account the effect of the time lag between samples. In dynamic speckle signals, the autocorrelation function $R_k$, is usually a monotonically decreasing function expressed as

$$R_k = \sum_{n=1}^{N} x_{n+k} x_n \quad \text{where } k \geq 0,\ k=\{0,...,N\} \quad (3)$$

A normalized version is typically employed, which is the result of dividing (3) by $R_0$ (at 0 lag). Hence, the descriptor computes the lag of the $R_k$ at its half maximum (*FWHMA*) [13].

3.2 Descriptors based on frequency analysis

The information of a time-varying signal is usually better interpreted from the frequency point of view. The Fourier transform of the autocorrelation function is the Power Spectral Density (PSD), which can be also employed to describe signal activity.

Fig 1 shows two intensity series of simulated speckle patterns at low and high velocities, together with their respective Power Spectral Density estimation. In order to reduce artifacts in the Fourier transform due to the finite-length of the discrete signal, the PSD is computed using the Bartlett-Welch estimator [14].

The multispectral nature of the speckle signal and its analysis in the frequency domain give rise various descriptors of interest.

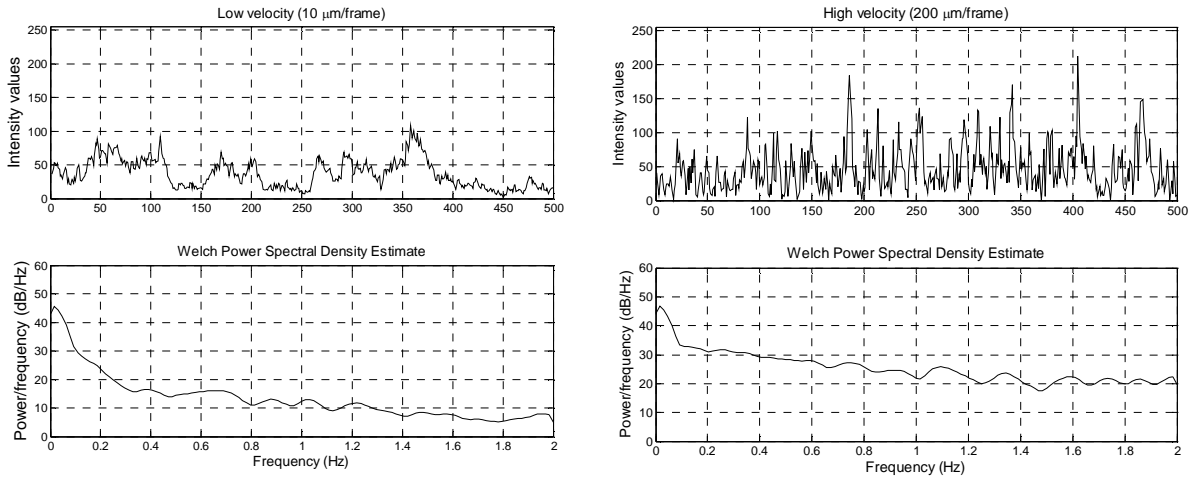

Figure 1. intensity series of simulated speckle patterns at low and high velocities

*3.2.1 Energy of spectral band (ESB)*

Under the hypothesis of the usefulness of the information provided by the spectrum, a descriptor is proposed based on the energy of the filtered signal. Sometimes, filtering the pattern intensity in the frequency domain helps to find a particular band that can characterize the boiling speckle. Infinite Impulse Response filters (IIR) are usually preferred to Finite Impulse Response (FIR) ones, given that a sharper transition region roll-off than FIR filter of the same order can be achieved.

The descriptor is then computed as the energy of the previously filtered signal $X$ [14].

$$ESB = \frac{1}{N}\sum_{n=1}^{N} x_n^2 \qquad (4)$$

There is no a priori method to know in which frequency band is the "useful" information of the signal. Although the *ESB* is also an empirical descriptor, an initial bank of filters can be applied to detect the bands of interest [15]. The *ESB* is the easiest analysis of the phenomena that originated the speckle signal. Finally, the *HLR* descriptor (see next section) can also be applied to relate two frequency bands.

For the case of low frequency bands, a prior subtraction of the mean value of the signal ($\langle X \rangle$) is required. This descriptor has turned out to be also effective to characterize some processes under non-uniform illumination, where each signal value is pre-processed with the following expression:

$$\tilde{x}_i = (x_i - \langle X \rangle)/\langle X \rangle \qquad (5)$$

*3.2.2 High to Low Ratio (HLR) of the Power Spectral Density*

Fujii et al [16] proposed the high to low frequency ratio (*HLR*) of two empirically selected frequencies.

$$HLR = \frac{s_H}{s_L} \qquad (6)$$

where $s_H$ and $s_L$ are the values of the PSD at a high and low frequency respectively.

Generally considered $f_{max}$ as the maximum frequency of the PSD decomposition, $s_L$ is computed within the interval from $f=0$ to $0.25\ f_{max}$ and $s_H$ is computed from $0.25\ f_{max}$ to $f_{max}$.

*3.2.3 Mean frequency (MF)*

Another descriptor of the PSD is the mean frequency (*MF*) introduced by Aizu and Asakura [17], defined as

$$MF = \frac{\sum_{k=1}^{N_f} f_k s_k}{\sum_{k=1}^{N_f} s_k} \qquad (7)$$

where $s_k$ is the $k$-th component of the PSD and $f_k$ its frequency value and $N_f$ is equal to the total number of frequency components.

*3.2.4 Cutoff frequency (CF)*

This descriptor computes the frequency at the half of the PSD maximum, hence this procedure is analogous to the *FWHMA*, but applied now to the PSD instead of the autocorrelation [17]

*3.2.5 Shannon Entropy (SE) of the Power Spectrum Density*

An alternative descriptor of the PSD is the entropy introduced originally by Shannon as a measure of the signal "disorder" [18]. The spectral entropy value is obtained multiplying the PSD estimate [$Pf(X)$] by its natural logarithm [$log_e Pf(X)$], summed over all $f$ frequency components. Thereafter, the entropy value is normalized to range between 1 (maximum irregularity) and 0 (complete regularity) [19]. The value is also divided by the maximum of the spectrum entropy of $N_f$ samples ($\log N_f$) where $N_f$ is equal to the total number of frequency components and is mathematically expressed as

$$SE = \frac{-\sum_{f=1}^{N_f} \frac{s_f}{S} \log\left(\frac{s_f}{S}\right)}{\log(N_f)} \qquad (8)$$

where $s_f$ is the $f$-th element of the power density spectrum vector **s** of length $N_f$, and $S$ is

$$S = \sum_f s_f \qquad (9)$$

*3.2.6 Discrete Wavelet Transform Entropy (DWTE)*

Instead of using *PSD*, the Shannon Entropy can also be applied to the decomposition of the signal using the *DWTE* [20][21]. *DWTE* makes no assumptions about signal stationary feature [10], so it provides a useful tool for the frequency analysis considering the temporal location [22]. This feature makes it possible to compute the time evolution of its entropy as a measure of the variation of the complexity or disorder of the time series. The wavelet transform (WT) is the representation of a signal *X* by means of its inner products with a set of basic wavelet functions.

A wavelet family $\psi_{a,b}$ is the set of functions generated by dilations and translations of a unique admissible mother wavelet $\psi(t)$

$$\psi_{a,b}(t) = |a|^{-1/2} \psi\left(\frac{t-b}{a}\right) \qquad (10)$$

where $a, b \in \mathbb{R}$, $a \neq 0$ are the scale and translation parameters respectively and $t$ is time. The wavelet becomes narrower accordingly as *a* decreases. Wavelet transform provides a tool for simultaneously observing a time series at a full range of different scales *a*, while retaining the time dimension of the original data. Multi resolution analysis theory shows that no information is lost if the continuous wavelets coefficients are sampled at a sparse set of points in the scale-time plane known as the dyadic grid. This grid leads to the *DWTE*, where the scale parameter is $a_j=2^{-j}$ and the translation $b_{j,k}=2^{-j}k$, with $j, k \in \mathbb{Z}$.

If the signal is assumed to be given by a set of sampled values corresponding to an uniform time grid, carrying out the decomposition over all resolution levels $M=\log_2(N)$, the wavelet expansion will be:

$$S(t) = \sum_{j=0}^{\infty} \sum_k C_j(k) \psi_{j,k}(t) = \sum_{j=0}^{\infty} r_j(t) \qquad (11)$$

where $C_j(k)$ can be interpreted as the local residual errors between successive signal approximations at scales *j* and *j+1* and $r_j(t)$ is the residual signal at scale *j*. The energy at each resolution level $j=-1,...,-M$. ($M=\log_2(N)$), will be the energy of the detail signal, at a given time window.

In order to study the time evolution of the speckle pattern, signal *X* is divided into temporal windows *i* of length *L* The following expression is used to obtain the mean wavelet energy of the detail signal *j* at each time window *i*

$$E_j^{(i)} = \sum_{k=0}^{(L/2^j)-1} |C_{k,j,i}|^2 \qquad (12)$$

with ($i=1,...,NT$, with $NT=N/L$, The total energy at interval *i* can be obtained by:

$$E_{total}^{(i)} = \sum_{j<0} E_j^{(i)} \qquad (13)$$

The signal window *i* relative wavelet energy will be given by:

$$p_j^{(i)} = \frac{E_j^{(i)}}{E_{total}^{(i)}} \qquad (14)$$

The following expression is used to evaluate the window *i* Shannon entropy [21]. The obtained value is assigned to the central window point, normalized with the log of the maximum level of decomposition, to obtain entropy values in the [0,1] interval.

$$DWTE = \frac{-\sum_{j<0} p_j \cdot \log[p_j]}{\log M} \qquad (15)$$

Consequently if the embedded behavior changes of the intensity series can be characterized by the time evolving entropy value, this parameter can be considered as a descriptor of the dynamic biospeckle. This descriptor has been successfully applied to the assessment of time varying phenomena like the paint drying of acrylic enamel [10]

3.3 Descriptors based on Time Domain Analysis

The following descriptors are based on the time domain calculations; hence, they strongly depend on the number of signal samples. In order to avoid this effect, a normalization division by the quantity of samples (*N*) was included in their mathematical expressions.

*3.3.1 Averaged Differences (AD)*

Due to the non-linearity of the speckle phenomenon, the mean intensity value is not suitable to weight the standard deviation. An alternative approach was introduced by Fujii et al [23], called *Averaged Differences* and also known as Fujii's descriptor. Here the difference between contiguous samples is weighted by the local average by the following expression:

$$AD = \sum_{n=2}^{N} \frac{|x_n - x_{n-1}|}{|x_n + x_{n-1}|} / N \qquad (16)$$

where |.| indicates absolute value.

*3.3.2 Generalized Differences (GD)*

Although the *AD* descriptor is very suitable for many applications, it suffers of two disadvantages: it is very sensible to the noise in regions of low intensity values and it is not suitable to detect slow varying speckle signals. In order to adapt it to these situations, a descriptor called Generalized Differences was presented by Arizaga et al [24], where intensity variations in different time scales are taken into account using the following expression:

$$GD = \left(\sum_n \sum_l |x_n - x_{n+l}|\right) / N \qquad (17)$$

where *n* and *l* are indices spanning all the possible numbers of the registered images. As every $x_n$ value is subtracted from every other value in *X*, the result does not depend on the sequence. This descriptor is thus very sensitive to the number of samples.

*3.3.3 Weighted Generalized Differences (WGD)*

An additional parameter ***p*** was later added to the GD descriptor with the aim of controlling its sensitivity. The elements of this vector allow giving different weights according to the gap of each subtraction.

$$WGD = \left(\sum_{n=1}^{N-p} \sum_{l=1}^{p} |x_n - x_{n+l}| p_l\right) / N \qquad (18)$$

The resulting descriptor, called Weighted Generalized Differences (*WGD*), was found useful to detect different types of activities. However, the selection of parameter *p* is highly empirical and limits its application from the practical point of view.

*3.3.4 Subtraction Average of consecutive intensities (SA)*

One of the simplest descriptor is the Subtraction Average (*SA*) of two consecutive elements of the *X* time speckle pattern [3].

$$SA = \sum_{n=1}^{N-1} |x_n - x_{n+1}| / N - 1 \qquad (19)$$

*3.3.5 Fuzzy Granularity (FG)*

Another novel descriptor is the fuzzy granularity (*FG*) which quantifies the time intensity variations [25]. The discrete signal is transformed into a set of fuzzy granules (*G*) characterized by three levels of intensity (*k*), *light, medium* and *dark*. The intensity levels are range values overlapped in the sense of fuzzy sets [26].

$$FG = \sum_{k=1}^{3} |suc_{n,k}(G(x_n,k))| / N \quad \text{with} \quad n = 2,3,...,N \qquad (20)$$

where

$$suc_{nk} = \begin{cases} 1 & \text{if } G(x_{n-1},k) \text{ is true and } G(x_n,k) \text{ is false} \\ 0 & \text{in other case} \end{cases} \quad (21)$$

Indicates the ending of a succession of the same level of intensity, and |.| indicates here cardinality, i.e., *FG* is the number of granules registered in *N* samples. The fuzzy set parameters are obtained from the intensity histogram of a speckle pattern; tuning this characteristic to a particular sample acquisition.

## 4. Results and discussion

4.1 controlled simulations

Sets of images with different controlled activities were simulated by using different speeds of the moving diffuser and applied as input to the different descriptors. Velocities from 1 to 200 *μm/frame* in steps of 2 were considered, and a set of 2000 samples of 512x512 pixels were obtained for each velocity.

The speckle activity was estimated by the different descriptors. Mean results corresponding to 512 time series (*X*) of 50, 200, 500, 1000 and 2000 samples are plotted against the simulated speed of the diffuser in micrometers per frame (Fig. 1 to 14). Then, in Table 1 are shown the linear correlation coefficient, the variation coefficient, the computing time for each descriptor with different samples quantity, and the relative difference among the results obtained with different samples number.

Figure 2 shows the *SD* as a function of the speed of the diffuser. In all the cases the descriptor was asymptotic to the same value. This descriptor can be useful only to describe low speeds and requires only a small number of samples. Results of *TC* descriptor (Fig. 3) are quite similar to *SD*, showing less noisy tendency than the *SD*.

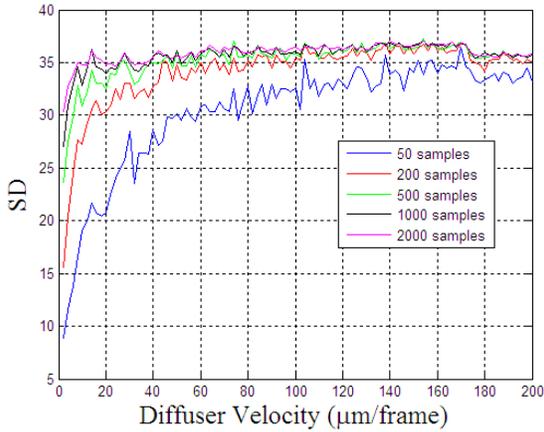

Figure 2: Standard Deviation descriptor (*SD*)

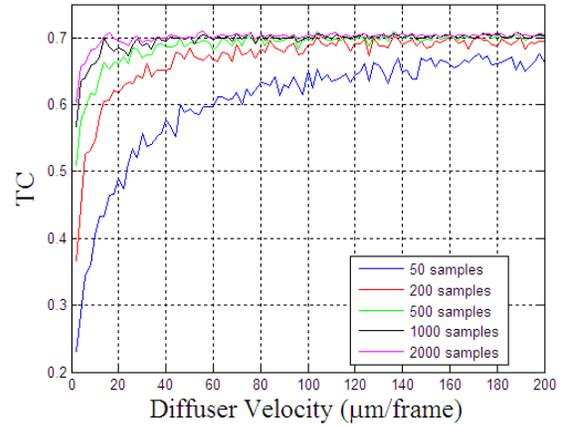

Figure 3: Temporal Contrast descriptor (*TC*)

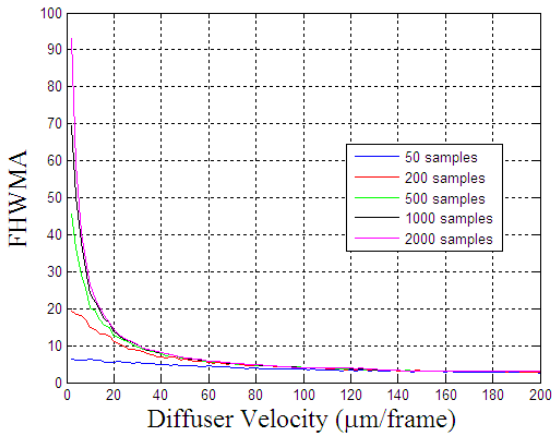

Figure 4: Full Width Half Maximum of the Autocorrelation

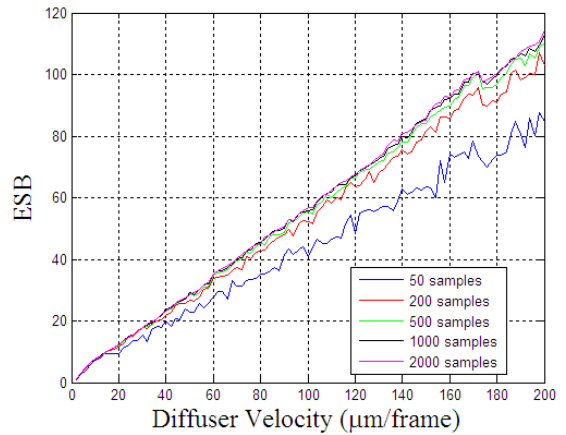

Figure 5: Energy of spectral bands descriptor (*ESB*)

(*FHWMA*)

Figure 4 shows that *FWHM* can distinguish speeds up to 60 *μm/frame* diffuser in our experiment.

The Energy of bandpass filter (0.25 Hz to 5Hz) is exhibited in Fig. 5. The losses specification in each of this filters are the following: maximum tolerated at the pass band 1db and the minimum losses required at the reject band is the 40db. Elliptic approximation (Cauer) filters were used because, in spite of their higher implementation complexity, they exhibit a more selective frequency response than Butterworth solution.

Fig. 5 to Fig. 10 shows Frequency descriptors results: *ESB, HLR, MF, CF, SE, DWTE* (with Daubechies db8). It is observed a high linearity behavior with the *HLR* (Fig. 6) and the *MF* (Fi.g 7) descriptors using more than 200 samples.

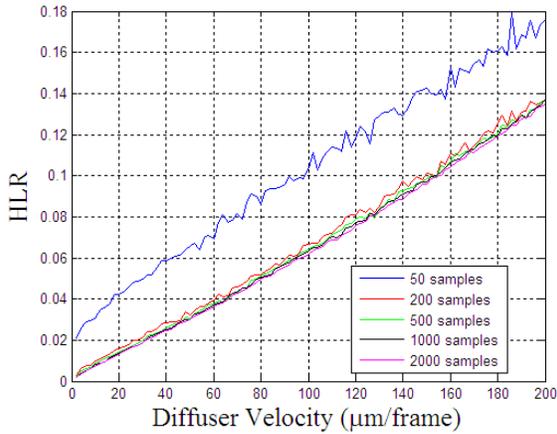

Figure 6: High-Low Ratio descriptor (*HLR*)

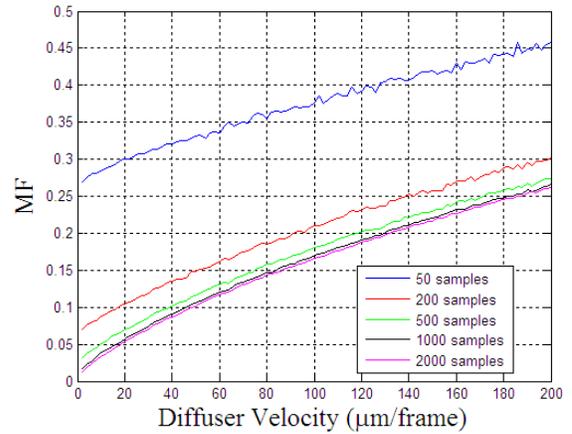

Figure 7: Mean Frequency descriptor (*MF*)

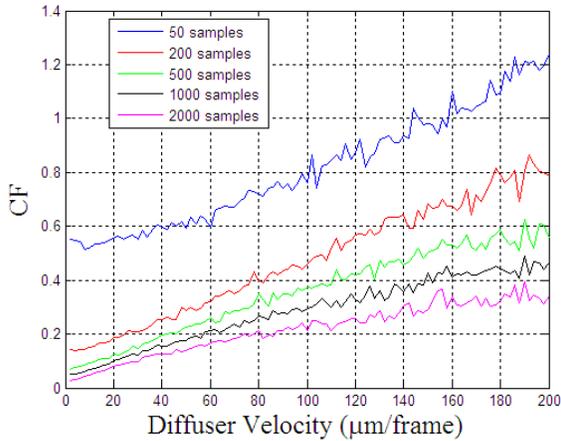

Figure 8: Cut-off frequency descriptor (*CF*)

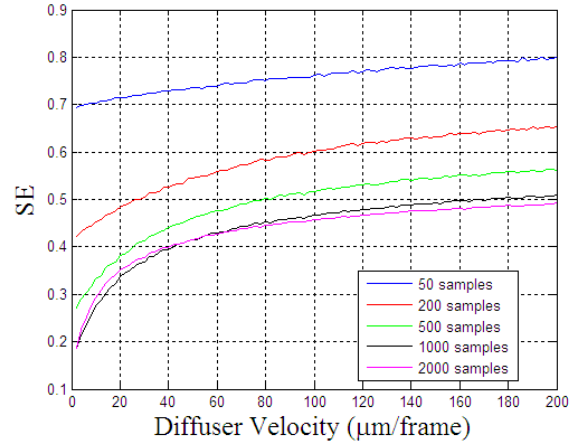

Figure 9: Shannon Entropy descriptor (*SE*)

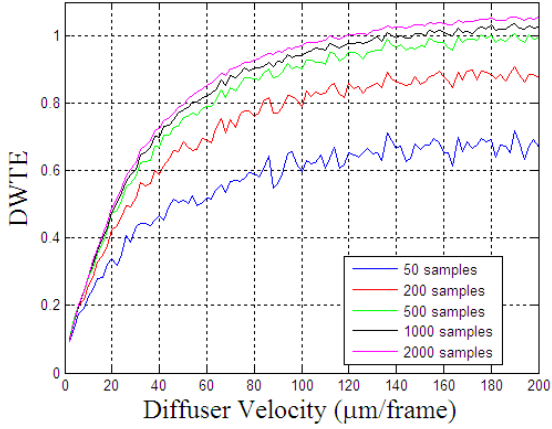
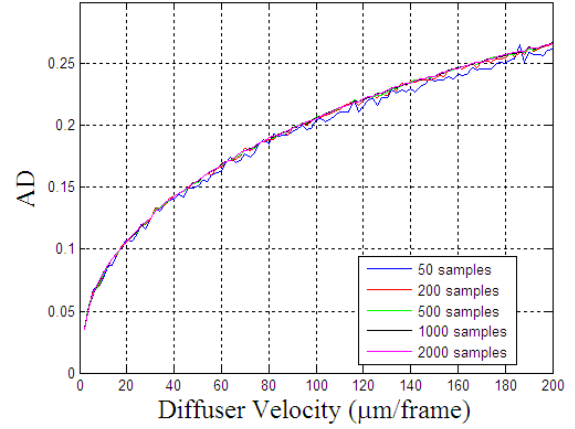

Figure 10: Discrete Wavelet Transform Entropy (*DWTE*)

Figure 11: Average Differences descriptor (*AD*)

The dependency of the frequency descriptors values with the samples quantity is highly relevant as observed in: *MF, CF, SE* and *DWTE*. Specifically in *HLR* a great disparity is observed for 50 samples.

Figure 11 shows the *AD* as a function of the diffuser speed. It can be seen that the trends show no appreciable noise. An advantage of this descriptor is its ability to differentiate a far wider range of activities.

Figure 12 shows the *GD* as a constant function of the diffuser speed; hence it is not appropriate to characterize the activity of pure boiling patterns generated by the displacement of a diffuser. Its bad behavior is overcome with the *WGD* descriptor Figure 13.

The *FG* fuzzy granularity (fig. 15) is comparable with the *SA* (fig. 14), the *AD* (fig.11) and *WGD* using *p=5* (fig.13).

The descriptors that exhibit less dependency to the number of samples are AD, SA and FG, followed by ESB and HLR.

In order to quantitatively assess the linearity of the descriptor in relation to the speed of the diffuser, the linear correlation coefficient is calculated. The Pearson's linear correlation coefficient shows a statistical relationship between the descriptor and the diffuser velocity values. It is defined as

$$r_{vd} = \frac{\sum_{i=1}^{L}(v_i - \langle v \rangle)((d_i - \langle d \rangle)}{(L-1)\sigma_v \sigma_d} \qquad (21)$$

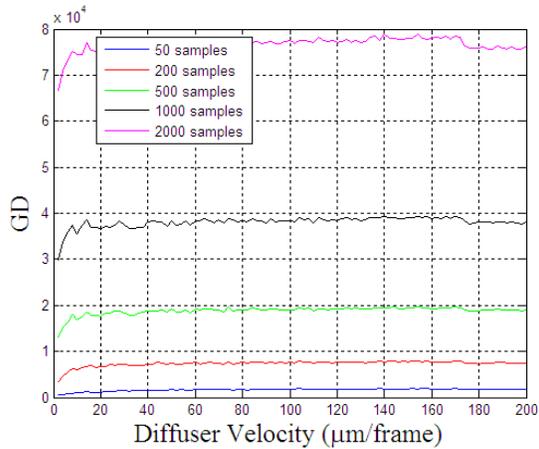
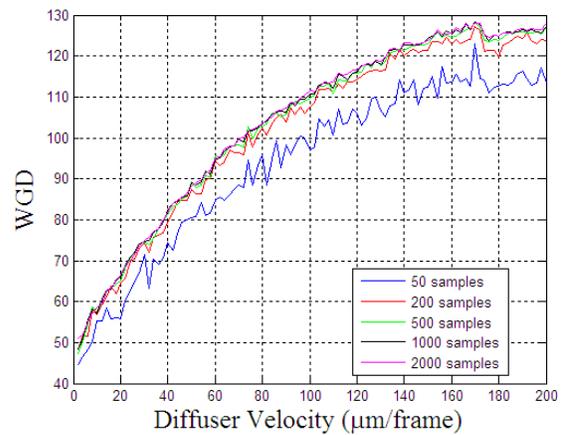

Figure 12 : Generalized Differences descriptor (*GD*)

Figure 13 : Weighted Generalized Differences descriptor (*WGD*)

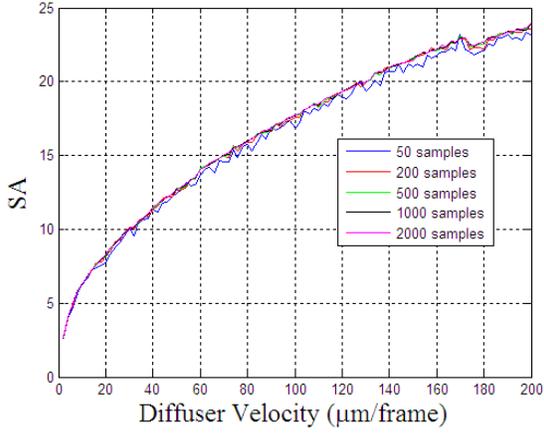 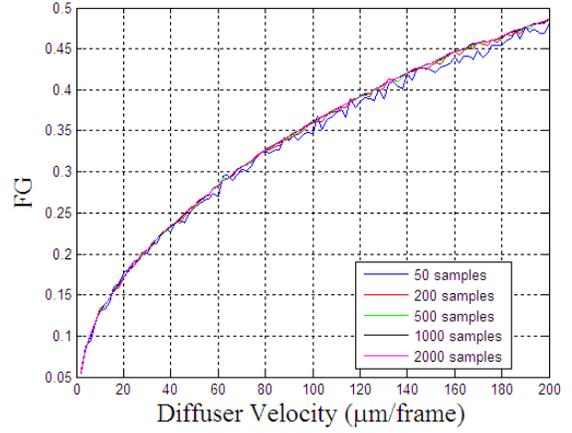

Figure 14 : Subtraction Average descriptor (*SA*)      Figure 15: Fuzzy Granular descriptor (*FG*)

where $v$ is the diffuser velocity value, $d$ is the descriptor value, $L$ is the number of considered velocities, $\sigma$ are the standard deviations and, $\langle v \rangle$ and $\langle d \rangle$ stand for v and d means respectively.

As an estimation of the robustness of the descriptors the coefficient of variation ($C_d$) is proposed, which shows the extent of variability in relation to the mean of the 512 values (one descriptor for each of the 512 time series). It is defined as the ratio of the standard deviation $\sigma_d$ to the mean $\langle d \rangle$ of the descriptor variable $d$:

$$C_d = \frac{\sigma_d}{\langle d \rangle} \qquad (22)$$

The sensitivity of the descriptor, $s_{di}$, is the relative difference between the value obtained using $i$ samples ($d_i$) and 2000 samples ($d_{2000}$)

$$s_{di} = \frac{|d_i - d_{2000}|}{d_{2000}} \qquad (23)$$

The computing time (*CT*) is the average of the elapsed time in processing each series in seconds.

In Table 1 are shown for each descriptor the variation coefficient $C_d$, the linear correlation coefficient $r_{vd}$, and computing time applied to different samples quantity.

| DESC | Correlation coefficient $r_{vd}$ | | | | | Variation coefficient $C_d$ | | | | |
|---|---|---|---|---|---|---|---|---|---|---|
| | 50 | 200 | 500 | 1000 | 2000 | 50 | 200 | 500 | 1000 | 2000 |
| SD | 0.794 | 0.650 | 0.601 | 0.545 | 0.537 | 0.318 | 0.190 | 0.131 | 0.095 | 0.069 |
| TC | 0.792 | 0.648 | 0.599 | 0.524 | 0.437 | 0.210 | 0.124 | 0.084 | 0.062 | 0.045 |
| FWHMA | 0.953 | 0.794 | 0.664 | 0.592 | 0.544 | 0.260 | 0.202 | 0.148 | 0.115 | 0.088 |
| ESB | 0.996 | 0.998 | 0.999 | 0.999 | 0.999 | 0.579 | 0.285 | 0.185 | 0.135 | 0.103 |
| HLR | 0.997 | 0.999 | 0.999 | 0.999 | 0.999 | 0.308 | 0.245 | 0.174 | 0.119 | 0.087 |
| MF | 0.995 | 0.994 | 0.993 | 0.992 | 0.992 | 0.090 | 0.097 | 0.086 | 0.066 | 0.050 |
| CF | 0.984 | 0.994 | 0.988 | 0.986 | 0.975 | 0.354 | 0.397 | 0.474 | 0.500 | 0.631 |
| SE | 0.986 | 0.948 | 0.904 | 0.875 | 0.852 | 0.024 | 0.034 | 0.040 | 0.036 | 0.028 |
| DWTE | 0.862 | 0.858 | 0.854 | 0.848 | 0.845 | 0.398 | 0.212 | 0.126 | 0.084 | 0.055 |
| AD | 0.963 | 0.964 | 0.964 | 0.964 | 0.964 | 0.182 | 0.097 | 0.062 | 0.045 | 0.032 |
| GD | 0.770 | 0.611 | 0.570 | 0.519 | 0.527 | 0.309 | 0.180 | 0.122 | 0.088 | 0.064 |
| WGD | 0.946 | 0.947 | 0.949 | 0.949 | 0.949 | 0.282 | 0.149 | 0.096 | 0.069 | 0.051 |

| | | | | | | | | | |
|---|---|---|---|---|---|---|---|---|---|
| SA | 0.974 | 0.974 | 0.975 | 0.975 | 0.975 | 0.201 | 0.109 | 0.072 | 0.054 | 0.043 |
| FG | 0.974 | 0.975 | 0.975 | 0.975 | 0.976 | 0.209 | 0.108 | 0.070 | 0.050 | 0.037 |

Table 1. Correlation and Variation coefficient

The ability of the descriptor to identify the dynamics of the process is evaluated with the correlation coefficient $r_{vd}$, calculated with its value and the diffuser velocity. High coefficients indicate a linear relationship with the simulated velocity. As is observed in Table 1 this condition is fulfilled in both temporal (*AD, WGD, SA, FG*) and frequency descriptors (*ESB, HLR, MF, CF, SE, DWTE*), though not in the statistics ones (*SD, TC* and *FWHMA*) neither in the GD.

A low variation coefficient is expected when defining a descriptor; this fact implies high reliability in its use. In Table 1 $C_d$, goes down with the increase of the number of samples, with exception of *CF*. *CF* shows the higher $C_d$, which makes it rather unreliable. The shadowed cells shows the best abilities in both features ($r_{vd} \geq 0.95$ and $C_d < 0.1$).

In Table 2 are shown the Sensitivity and the Computing Time average. The sensitivity $S_{di}$ shows the capability of the descriptor *d* to get a good performance still using a low quantity of samples *i*. The shadowed cells shows the best performances in both features ($S_{di}$ and $CT < 0.1$).

Table 3 displays comparative graphs of the four evaluated features:

| DESC | Sensitivity $S_{di}$ | | | | Computing time average *CT* [s] | | | | |
|---|---|---|---|---|---|---|---|---|---|
| | 50 | 200 | 500 | 1000 | 50 | 200 | 500 | 1000 | 2000 |
| SD | 0.156 | 0.051 | 0.021 | 0.009 | 0.001 | 0.002 | 0.004 | 0.011 | 0.026 |
| TC | 0.139 | 0.047 | 0.018 | 0.007 | 0.001 | 0.002 | 0.005 | 0.013 | 0.030 |
| FWHMA | 0.234 | 0.095 | 0.044 | 0.018 | 0.115 | 0.145 | 0.196 | 0.304 | 0.588 |
| ESB | 0.217 | 0.062 | 0.023 | 0.010 | 0.485 | 0.499 | 0.536 | 0.616 | 0.757 |
| HLR | 1.096 | 0.102 | 0.046 | 0.021 | 1.179 | 1.305 | 1.190 | 1.334 | 1.429 |
| MF | 2.236 | 0.468 | 0.150 | 0.037 | 1.181 | 1.304 | 1.188 | 1.330 | 1.424 |
| CF | 3.578 | 1.258 | 0.651 | 0.312 | 1.188 | 1.315 | 1.200 | 1.343 | 1.440 |
| SE | 0.777 | 0.351 | 0.132 | 0.026 | 1.187 | 1.310 | 1.195 | 1.336 | 1.433 |
| DWTE | 0.345 | 0.156 | 0.059 | 0.026 | 0.695 | 0.697 | 0.714 | 0.767 | 0.866 |
| AD | 0.021 | 0.008 | 0.004 | 0.003 | 0.001 | 0.004 | 0.013 | 0.033 | 0.068 |
| GD | 0.978 | 0.904 | 0.754 | 0.503 | 0.007 | 0.155 | 1.310 | 8.441 | 35.355 |
| WGD | 0.977 | 0.902 | 0.752 | 0.501 | 0.001 | 0.002 | 0.005 | 0.013 | 0.054 |
| SA | 0.023 | 0.008 | 0.005 | 0.004 | 0.001 | 0.002 | 0.005 | 0.014 | 0.028 |
| FG | 0.021 | 0.007 | 0.004 | 0.002 | 0.006 | 0.023 | 0.057 | 0.113 | 0.225 |

Table 2. Sensitivity and Computing Time average

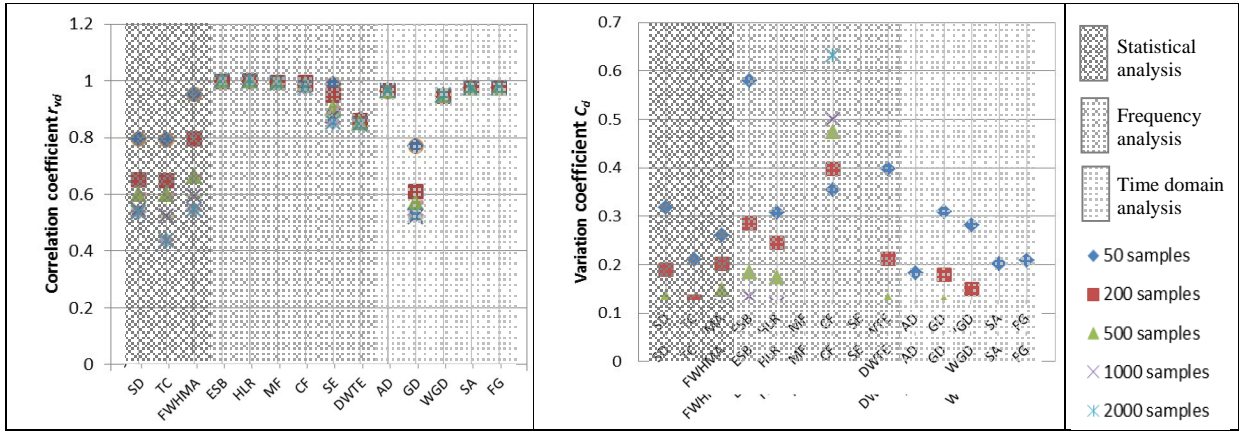

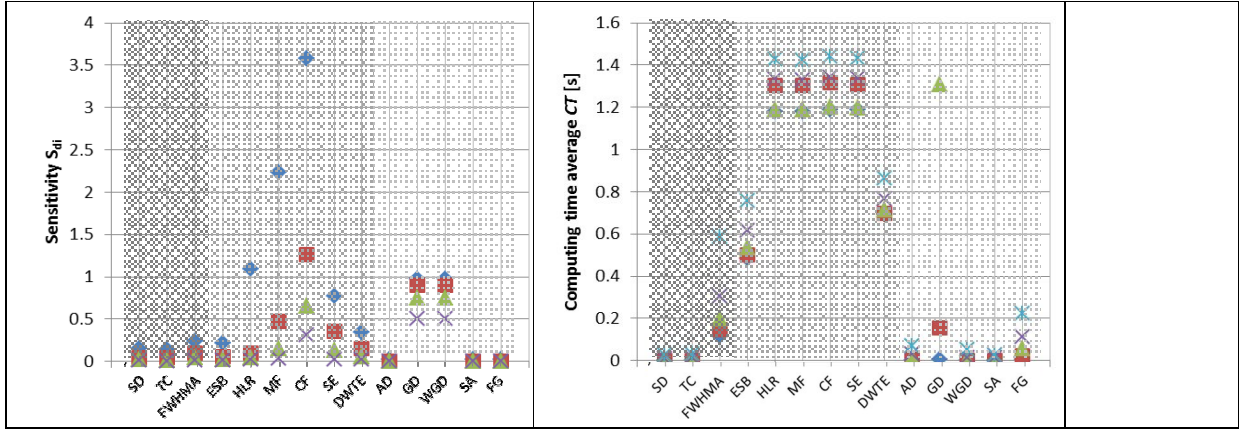

Table 3. Comparative graphs of the four evaluated features

4.2 Actual experiment: bruise-apple

In order to assess the robustness of the descriptor to the intensity variations in the area of a sample due to its topological characteristics, we evaluated the behavior of the descriptors in actual and complex experiments where non-visible a bruise in apples was detected [10].

The goal was the recognition of the region bruised amid the healthy one. To compare their performance the Receiver Operator Characteristics (ROC) was used [27], and the area under the curve (AUC) was computed as a quality index. The descriptors were obtained using two different ensembles: of 500 and 50 images each. As is widely known, the AUC higher value the better the classifier.

In Table 4 the AUC values for each descriptor are shown in descending order. Using 500 samples the descriptor with the best performance with the higher AUC is the *FG* with 0.954, followed by the frequency descriptor *MF, HLR, ES, DWTE, CF* and *ESB* with values above 0.75. Other descriptors show an inadmissible performance with values below 0.5, in consequence they are not advisable to apply on samples with inhomogeneous spatial intensity. When a set of 50 images was used to compute the descriptors, no majors differences are shown in the FG performance, followed by the frequency ones. Nevertheless the *TC* and the *FWHMA*, both based on statistical analysis, enhance their performance in relation with that shown with 500 samples.

|            | AUC         |            |
|------------|-------------|------------|
| Descriptor | 500 samples | 50 samples |
| FG         | 0.954       | 0.940      |
| MF         | 0.936       | 0.902      |
| HLR        | 0.926       | 0.910      |
| ES         | 0.896       | 0.895      |
| DWTE       | 0.887       | 0.692      |
| CF         | 0.862       | 0.830      |
| ESB        | 0.754       | 0.651      |
| TC         | 0.412       | 0.718      |
| FWHMA      | 0.276       | 0.667      |
| AD         | 0.276       | 0.276      |
| WGD        | 0.020       | 0.261      |
| SA         | 0.019       | 0.410      |
| SD         | 0.026       | 0.260      |

Table 4. Comparative graphs of the four evaluated features

## Conclusions

We have evaluated the performance of several biospeckle descriptors by applying them to controlled numerical simulations and also to a real experiment with non-visible bruises in apples. The performance evaluation with the numerical simulation of a controlled pure boiling speckle experiment was shown on plots of the value of the descriptor as a function of the diffuser simulated speed. The effect of changing the length of the sample, its convergence, linearity and standard deviation were assessed.

It should be noted that most of the descriptors need to have a priori the total set of samples, e.g.: those based in frequency and statistics meanwhile those as AD, SA and FG could be computed as new samples appear. This latter feature makes them candidates to be embedded in real time processes.

Thus, in the simulated experiment the set of descriptors based on statistical and GD do not have an acceptable coefficient of correlation. The coefficient of variation is acceptable for most experiments descriptors for over 500 samples except CF whose variability is high relative to the average. The descriptors based on frequency analysis as MF, CF, SE, and DWTE and also GD and WGD show high sensitivity to the number of samples used. The computation time is high for descriptors based on frequency analysis and also GD.

The result of the actual experiment shows that FG exhibits the best performance, followed by those based on frequency analysis, demonstrating that are robust to spatial changes of intensity within the sample, in agreement with the analysis of correlation in the simulations.

## Acknowledgements

This work was supported by CCT La Plata Consejo Nacional de Investigaciones Científicas y Técnicas (CONICET). Comisión de Investigaciones Científicas de la Provincia de Buenos Aires. by Facultad de Ingenieria. University of La Plata and by a Grant PICT 2008-1430. Agencia Nacional de Promoción de la Ciencia y la Técnica. Argentina.